# An "anti-system" ontology of quantum physics, as derived from two Einstein's conceptions of physical theories

## Quantum physics as a theory of general relativity of experimental context

**Thierry Batard**
*thierry.batard@outlook.com*

In glaring contrast to its indisputable century-old experimental success, the ultimate objects and meaning of quantum physics remain a matter of vigorous debate among physicists and philosophers of science. This article attempts to shed new light on the debate. It relies upon two comments by Albert Einstein on his general approach to physical theories. I draw their consequences for the definition of a physical theory's ontology, and next for the ontology of quantum physics—i.e. what it may ultimately be *about*. The quantum ontology thus derived appears to be strictly limited to evolving experimental contexts and instantaneous measurement outcomes, which are to be understood, respectively, as mere *potential* measurement outcomes and *actual* ones. The notions of material body in particular and physical system in general are absent from this ontology, hence the vanishing of Schrödinger's cat and EPR paradoxes, as well as of the quantum measurement problem. Apart from its ability to clear up well-known conundrums, this ontology reveals what quantum physics may fundamentally be—i.e. its possible ultimate *meaning*—namely a theory of general relativity of experimental context. On this basis, I conclude with a new conception of objectivity in the sciences of nature.

## 1  Introduction

Quantum physics is a theory whose "… experimental predictions… have always been right" (d'Espagnat 2012)[1], "… a theory that seems to account with precision for everything we can measure…" (Zurek 2002)[2]. Sixty years after Erwin Schrödinger's formulation of his famous quantum wave equation, another Nobel laureate in Physics speaks of an "enormous success"[3], and of the "sense of inevitability that quantum mechanics gives us" (Weinberg 1987).

We then might expect this theory to provide us with a clear understanding of the sensible world. Yet that is not the case at all. Indeed, "… we have great difficulties in saying *about what* it is and *what* it means, because that is precisely not self-evident" (Bitbol 1996, emphasis his). As a matter of fact, "it is well known that physicists, while they all agree about how to use Quantum Mechanics…, still all disagree about what it means, and even more about 'the real stuff' it describes: that is, its ontology" (Auffèves and Grangier 2016)[4]. Consequently, "… in spite of the variety of approaches developed with the aim of clarifying its content and improving the original formulation, quantum mechanics maintains a remarkable level of obscurity" (Rovelli 1996). Hence Feynman (1967) thinks he "… can safely say that nobody understands quantum mechanics."[5] Thus, it seems that with quantum physics "… science… has given up the quest for an intelligible world…"[6] In the eyes of the Fields medalist René Thom (2016), this makes quantum physics "… far and away the intellectual scandal…" of the twentieth century.

---

[1] "No experimental evidence is known which contradicts it", "… it agrees fully with experiment", Everett (1957) and Feynman (2006), respectively, acknowledge.
[2] "… We all agree that quantum mechanics works spectacularly well for every practical purpose…" (Fuchs et al. 2014).
[3] It is a "fantastic experimental success" (Greene 2004), a success that "… has long been proven…" (Bitbol 2019).
[4] This makes "… the understanding of quantum mechanics a vivid field of debates and interpretations, more than a hundred years after it was born", Auffèves and Grangier (2016) point out.

"Now that we are well into the 21st," Fuchs et al. (2014) comment, "… surely it is time… to dispel the murkiness that has obscured the foundations of the theory for too long." The present paper is part of an attempt in that direction. It relies upon two comments by Albert Einstein pertaining to his general approach to physical theories. I draw their consequences for the question "What quantum physics is *about*?", i.e. for its ontology. Once a quantum ontology has been so derived, I evaluate its consequences for our conception of the sensible world, for two paradoxes of quantum physics, as well as for the quantum measurement problem. After discussing the philosophical implications of this ontology, I attempt to answer the question "What is the *meaning* of quantum physics?", i.e. "How to make sense of its formalism?"

Besides Einstein's comments, the analysis of quantum theory by the French physicist and philosopher Michel Bitbol largely backs up this paper.

## 2 Two Einstein's general approaches to physical theories and their impact on the definition of "ontology"

Fig. 1 below reproduces a diagram from Albert Einstein intended to summarize his general conception of physical theories for his friend Maurice Solovine.

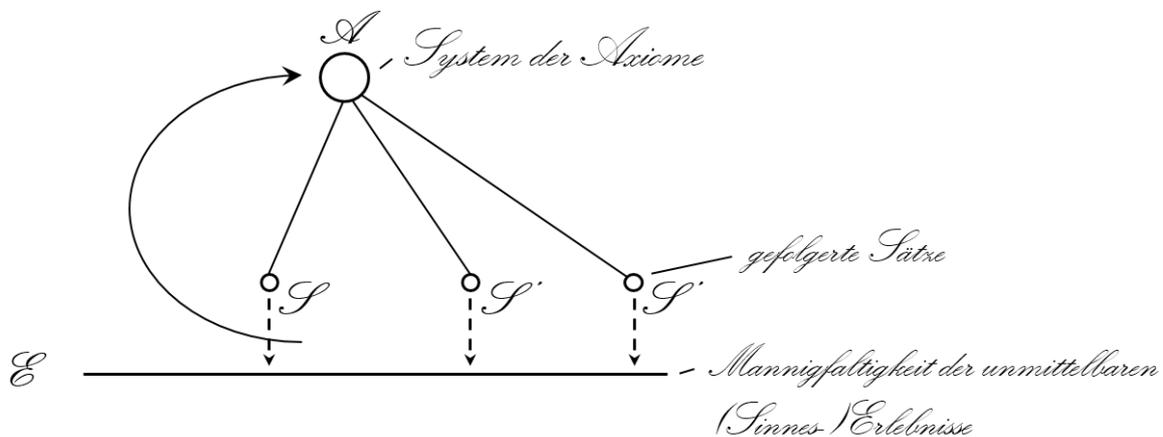

**Fig. 1**    Reproduction of a diagram from Albert Einstein (see Holton 1998 for the original one)

---

The German-born physicist explains his diagram to Solovine as follows:

"(1) The *E* (experiences[7]) are given to us.

"(2) *A* are the axioms[8], from which we draw consequences. Psychologically the *A* rest upon the *E* There exists, however, no logical path from the *E* to the *A* but only an intuitive (psychological) connection, which is always 'subject to revocation'.

"(3) From the *A* *by a logical route,* are deduced the particular assertions *S* which deductions[9] may lay claim to being correct[10].

"(4) The *S* are referred to the *E* (test against experience). This procedure, to be exact, also belongs to the extra-logical (intuitive) sphere, because the relation between concepts that appear in *S* and the experiences *E* are not of a logical nature."[11]

Furthermore, during a 1926 conversation with Werner Heisenberg, Albert Einstein claims that "only the theory decides what one can observe"[12]. In other words, what can be observed according to a physical theory should be determined *a posteriori,* that is, after the theory has been developed from axioms *A*[13]. It should not be determined before *(a priori),* with the aim of developing the theory on this basis, that is afterward—as Heisenberg thinks at the time of the discussion with Einstein[14]. Let us consider this statement together with Einstein's explanation of his above diagram. Then, what can be observed according to a physical theory are the *concepts*—hereafter referred to as *B*[15]—that appear in the deductions *S* of this theory and that can "be referred"[16] to experiences *E* given that the latter must be *immediate* sense experiences (hence the observability of the concepts *B*), as per Einstein's diagram. To take Einstein's example, "the notion 'dog'" *(der Begriff Hund)* can "correspond" *(entsprechen)* to the immediate experiences that one has while saying "I see a dog" (Feinberg 1998; Holton 1998). In the presence of such experiences, one can well say "I *observe* a dog"[17].

Hereafter, the word "ontology" will be used to designate the set of concepts *B* which, following the method described above, can be considered observable according to a physical theory, after the latter has been formalized. The word "ontology" is thus used in a non-*metaphysical* sense, as the concepts in question are related to sense experiences, and are consequently not "beyond[18] what concerns nature[19]", that is, beyond the *sensible* world.

# 3 Derivation of the ontology of quantum physics in light of Einstein's approaches to physical theories

That being said, what are the observable concepts $\mathscr{B}$ of quantum physics? What is its ontology? To answer those questions, let us consider its mathematical formalism ("the deductions $\mathscr{S}$" Einstein would say[20]). It will be assumed that this formalism can be reduced to the two following equations, that is, respectively, the Schrödinger equation (e.g. Zurek 2002) and the one corresponding to the Born rule (e.g. Bitbol 1996):

$$i\hbar \, \frac{\partial}{\partial t} |\psi\rangle = \hat{H} |\psi\rangle, \tag{1}$$

$$P(a_i, \psi) = |\langle a_i | \psi \rangle|^2. \tag{2}$$

The questions above now take the following form: what concepts $\mathscr{B}$ appearing in equations (1) and (2) can be referred to *immediate* sense experiences $\mathscr{E}$?

Obviously, it can neither be the imaginary unit $i$ (whose square is $-1$), nor the reduced Planck constant $\hbar$ (approximately $1.05 \times 10^{-34}$ joule-seconds).

In contrast, the concept of time represented by the "$t$" symbol can perfectly correspond to immediate sense experiences $\mathscr{E}$, namely, those one has when reading clocks, as precisely prescribed by Einstein for measuring time (Einstein 1905a).

In the same manner, the state vector $|\psi\rangle$, that characterizes an "experimental preparation" preceding a physical measurement[21], can be referred to the immediate sense experiences $\mathscr{E}$ of this experimental preparation, that is, those one has once the corresponding equipment (source, filter, …) is installed (Bitbol 1996). However, it must be stressed that the "$|\psi\rangle$" symbol does not at all designate this equipment, but the very concept of experimental preparation, understood as "an information potential… about the subsequent measurement results" (Bitbol 1996)[22]. As a matter of fact, the state vector $|\psi\rangle$ symbolizes—or expresses—the "predictive content" of this experimental preparation, the evolution of which is governed by equation (1) (Bitbol 1996)[23]. Therefore, "$|\psi\rangle$ does not refer… to objects, but to *situations*"[24], exactly in the same way that the "$t$" symbol does not refer to clocks but to the continuous variable called "time", although it is related to the experiences $\mathscr{E}$ one has in the presence of clocks.

---

[18] Late translation of the ancient Greek word "meta" (μετά).

[19] "Phusikos" (φυσικός) in ancient Greek.

[20] The formalism of quantum physics may not be deduced from a system of axioms as naturally as the formalism of Newton's physics or that of Einstein's theory of relativity. Nevertheless, one must rely upon the formalism of quantum physics in order to derive its ontology as per the method described in the previous section.

[21] "… It characterizes a *preparation* in its connections with any possible subsequent measurement" (Bitbol 1996, emphasis his).

[22] "… Theoretical symbols, such as the state vector, … aim… only at evaluating the probability of the discontinuous experimental events" (Bitbol 2019).

[23] "The operation of state preparation $P_\psi$ acts as a *noncognitive,* purely *physical* definition of an *infinite* set… of mere potentialities" (Mugur-Schächter 1994, emphasis hers), and the quantum state is merely "a probability list for all possible measurement outcomes" (Ma et al. 2012).

[24] As per a so-called hermeneutical interpretation of "the potentiality concept to which refers the vector $|\psi\rangle$ corresponding to an experimental preparation" (Bitbol 1996, emphasis added).

With regard to the scalar product $\langle a_i | \psi \rangle$, that gives the modulus of the projection of the state vector $| \psi \rangle$ onto the axis parallel to $| a_i \rangle$, it refers to the variable (which is named "*A*" for the values noted "$a_i$") that is measured following the experimental preparation characterized by $| \psi \rangle$, and that is determined by the apparatus used for this measurement (Bitbol 1996) and its configuration. The "$\langle a_i | \psi \rangle$" symbol, thus referring to the concept of "instrumental context" of a physical measurement (in the words of Bitbol 1996 and 1998), therefore corresponds to the immediate sense experiences $\mathscr{E}$ one has once the measuring apparatus is installed and configured for the intended measurement. It must be stressed again, however, that the "$\langle a_i | \psi \rangle$" symbol does not designate the apparatus and its configuration, but a restriction of all the possible measurement results that can be obtained following an experimental preparation, as characterized by the state vector $| \psi \rangle$.

The Hamiltonian $\hat{H}$, a mathematical operator called Hermitian, refers to the "intermediary circumstances between the preparation and the measurement", such as the "electric, magnetic, gravitational, static or variable fields" (Bitbol 1996). It then corresponds to the immediate sense experiences $\mathscr{E}$ commonly associated with the instruments or objects that define those intermediary circumstances. Once again, the concept $\mathscr{B}$ related to such experiences $\mathscr{E}$ is not the set of instruments or objects in question, but the experimental preparation characterized by $| \psi \rangle$ and considered *in its evolution,* as governed by equation (1), in which the Hamiltonian $\hat{H}$ appears.

In fact, the distinction between the concept of experimental preparation and that of instrumental context, as defined above, is rather artificial. Indeed, between the end of the preparation and the measurement, there is an evolution phase whose calculation is based on the state vector, according to "an extensive conception of the *preparation*"[25], but can just as well be based on the system of orthogonal axes representing the variable to be measured in the Hilbert space, according to "an extensive conception of the *measurement*" (Bitbol 1996, emphasis added)[26]. It is equally artificial to consider separately the concept of intermediary circumstances, since the latter have an influence on the evolution of the experimental preparation. That is why the three concepts, namely the experimental preparation, the instrumental context, and the intermediary circumstances, will be subsumed under the overall concept $\mathscr{B}$ of "experimental context"[27]. This overall concept will then be related to *all* corresponding immediate sense experiences $\mathscr{E}$ that is, the ones corresponding to "$| \psi \rangle$", "$\langle a_i | \psi \rangle$", and "$\hat{H}$" symbols. An experimental context can thus be defined as follows: an information potential (as symbolized by $| \psi \rangle$) about the outcome of a future measurement, which potential is constrained in its evolution by the intermediary circumstances (as symbolized by $\hat{H}$)[28], and is further constrained by the variable to be measured (as symbolized par $\langle a_i | \psi \rangle$)[29]. An experimental context corresponds to the "established experimental program", i.e. to the *whole* preparation of a physical *measurement*. That is why (a) it especially includes the instrumental context, and (b) the ontology presented here differs from a mere ontology of state vectors, a fortiori from the ontology of "a single 'universal state vector'" (Bitbol 1996).

---

[25] Cf. equation (1). As a matter of fact, this option corresponds to a so-called "Schrödinger representation" (Bitbol 1996).

[26] This option corresponds to the so-called "Heisenberg representation" (Bitbol 1996).

[27] In the words used by Michel Bitbol (e.g. 1996 and 2019). One might as well speak of "experimental conditions", "experimental circumstances" or "experimental situations" (Bitbol 1996).

[28] The absence of specific circumstances (fields, …) is still a circumstance.

[29] "… The indication of the type of used apparatus, represented in the formalism by an operator called 'observable'… *determines the range of possible events* whose probabilities are calculated" (Bitbol 1996, emphasis added). In other words, "… the ranges of values that quantum observables can take are related to the *types of equipments* used (devices for measuring position, momentum, spin components…)" (Bitbol 2019, emphasis his).

As for the "$a_i$" symbol considered in isolation, it simply corresponds to the concept $\mathscr{B}$ of measurement outcome—or measurement result or measurement event—i.e. the value taken at the time of measurement by the variable $A$ determined by the installed measuring apparatus and its configuration (Bitbol 1996). This concept of measurement outcome is related to the immediate sense experiences $\mathscr{E}$ one has while recording a physical measurement, e.g. while visualizing "a metric indication on the screen of a measurement instrument" (Bitbol 1996). Yet, the "$a_i$" symbol does not refer to this screen or to any "concrete" means of displaying a measurement result, but to the very concept of measurement outcome, such as a given number.

At last, the "$P(a_i, \psi)$" symbol merely designates a *probability:* the probability of obtaining result $a_i$ while measuring the variable $A$ following a preparation characterized by $|\psi\rangle$ and its evolution governed by equation (1). That is why the "$P(a_i, \psi)$" symbol, considered as a whole, cannot be referred to sense experiences $\mathscr{E}$[30].

Overall, the quantum ontology derived above clearly takes account of the necessary link—or the "umbilical cord"—between the formalism of a physical theory and the "physicist's lived, concrete experiences", namely "what she/he is experiencing while actively arranging her/his equipments and perceiving experimental events" (Bitbol 2019). With such an ontology, therefore, not only is the physicist immune against the temptation to "forget about" those experiences, but she/he is forced to fully assume them.

# 4 Some consequences of the derived quantum ontology

## 4.1 Consequence for the conception of the sensible world

According to the approach followed above, the ontology of quantum physics is a combination of (a) experimental contexts that—possibly—evolve with time and (b) measurement events occurring at given moments in time, this ontology being *strictly limited to those two concepts (a) and (b)*. An experimental context is then to be understood as a mere set of constrained possibilities of a future measurement outcome. Following Aristotelian terminology (see Burnyeat 2008), the quantum ontology derived above is limited to measurement outcomes (a) "in potentiality" (*dunamei* δυνάμει, in ancient Greek) and (b) "in actuality" (*energeia* ἐνεργείᾳ)*,* or—to be even more concise—to (a) "potential" and (b) "actual" measurement outcomes. In that spirit, Bitbol (2019) speaks respectively of (a) "a superposition state of an observable's eigenvalues, in which it can be considered that nothing has occurred except the protean potentiality to occur", and (b) "the actuality of a single value of the observable", hence "the plurality of the possibilities and the unicity of the actuality". Of note, Werner Heisenberg's interpretation of the quantum state vector already consists in "… introducing a quantitative version [of] Aristotle's notion of δύναμις (in Latin: *potentia*) to physics…" (Jaeger 2017; see also Heisenberg 1959 and 1962). And as he so rightly says, the so-called "… discontinuous 'reduction of wave-packets'… is… a consequence of the transition from the possible to the actual" (Heisenberg 1955). It corresponds to "the actualization of the potentia that occurs upon measurement", hence "the similarities between the actualization of quantum potentiality in measurement and actualization of δύναμις in Aristotle's philosophy", as "exhibited by Heisenberg" for whom "… the actualization of the potentialities taking place upon measurement is ontological rather than epistemic…" (Jaeger 2017). When a measurement has taken place, "… a certain one among the various possibilities has proved to be the real one…" (Heisenberg 1955) or, to be more exact, "the actual one".

---

[30] As with a classical probability calculation, the formalism of quantum physics is "… unable to tell by itself, both after and before random processes are achieved, *which* of the events occurs following certain antecedents" (Bitbol 1996, emphasis his).

Noticeably, *material bodies in particular and physical systems in general are absent from the quantum ontology derived in this paper.* That is why personally I never use the term "quantum mechanics". Indeed, the word "mechanics" denotes a branch of physics that deals with the behaviour of physical systems (Cleveland and Morris 2015) such as material bodies (Wigner 1967), which are the archetype of the physical systems. It is worth noting about quantum physics that "… its essential traits allow themselves to be grasped almost immediately when one *starts* to regard it as a mere predictive symbolism" and, to this end, to "… completely dispense with the concept of a physical system *on which* a measurement would be carried out"—against Paulette Destouches-Février's assertion that "the fundamental concept in physics is that of system" (Bitbol 1996, emphasis his). The approach followed above brings us back to the starting point of a "predictive symbolism", with the difference that quantum physics has now an ontology attached to it—should the concept of physical system be absent from this ontology. There is then no question of "restricting science to the 'meaningless empiricism' denounced by Einstein" (Bitbol 1996), as science has now something to tell us about the sensible world—while still leaving the instrumentalist fully satisfied, I believe (see section 5 below). Let us point out that, although Heisenberg contends that "… quantum mechanics has brought the concept of potentiality back into physical science" (Northrop 1962), he still refers to systems and their properties. He indeed views "… the state vector as the precise, mathematical encoding of this potential of the quantum system to possess any observed physical properties upon its measurement in the future" (Jaeger 2017), while I view it as the encoding of a mere potential future measurement outcome.

As a consequence of physical systems being excluded from the quantum ontology, there is no—longer—question of referring in quantum physics to atoms, electrons, quarks, and particles in general[31]. There is no question either of referring to laboratory instruments: sources, measuring apparatus, …[32] Regarding specifically the experimental context, instruments must be regarded as mere mental representations of it, which representations must only help to characterize its three constitutive elements, namely, the experimental preparation, the intermediary circumstances, and the type of measurement result that will be obtained[33]. Once this characterization is achieved, any reference to laboratory instruments must be scrupulously rejected, just as Wittgenstein's ladder must be thrown away after one has climbed up it (Wittgenstein 1921, § 6.54)[34]. Let us not forget that the state vector "$|\psi\rangle$ does not refer… to objects, but to situations" (see section 3 above). And the same can be said of "$\langle a_i|\psi\rangle$" and "$\hat{H}$" symbols, referring to elements that complete the characterization of those situations. In other words, the immediate sense experiences $\mathscr{E}$ underlying the mental representations of laboratory instruments refer to the experimental context *through* those representations. The overall concept $\mathscr{D}$ of experimental context is—so to speak—abstracted from the latter, the same way that the concept of time is abstracted from mental representations of clocks, or the concept of measurement outcome is abstracted from the representation of an apparatus screen (Fig. 2).

---

[31] "There are no quantum jumps, nor are there particles!" says the title of a quantum physics paper (Zeh 1993).

[32] Bitbol (1996) speaks of "system-equipment".

[33] Or "the range" it belongs to. See footnote 29 above.

[34] In that spirit, Bitbol (2019) notes that "the ladder of the ontology of things and facts must be thrown away in due course after it has helped reaching higher levels of knowledge…"

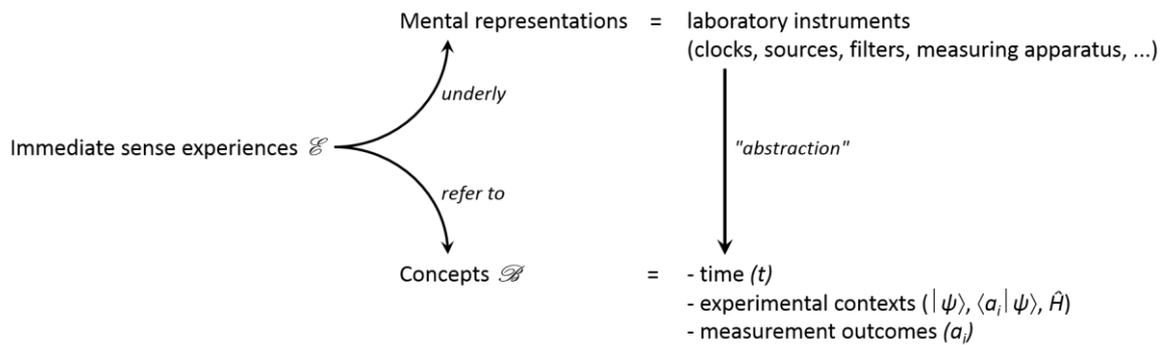

**Fig. 2** Immediate sense experiences, mental representations, and concepts, in connection with the quantum formalism

## 4.2 Consequence for Schrödinger's cat paradox

Since there is no question of referring to any physical system with the quantum ontology derived in this paper, there is of course no question of referring especially to the notion of dog (*der Begriff Hund;* see section 2 above) or that of cat. Hence the famous Schrödinger's cat paradox (Schrödinger 1935 and, e.g. Bitbol 1996 and 2019) is dissolved. With no possible reference to a cat, the question does not arise as to what the cat's state is between the end of the preparation of the "diabolical" thought experiment and the moment of the measurement which, following our naïve classical world view, consists of what we would call "the observation of the cat's biological state (dead or alive)". The immediate sense experiences usually related to the concept of cat—on the basis of which either a living cat or a dead cat is said to be observed—can now only relate to the concepts of experimental context and measurement outcome, as understood above. The concepts of dead cat and living cat must be replaced by the notion of a mere element of characterization of the experimental context (before measurement) and by the notion of a mere measurement event (at the moment when the latter occurs). Then, what happens between the end of the experimental preparation and the moment of the measurement is no longer an issue. Between the two moments, nothing happens except the evolution, as governed by the Schrödinger equation, of the experimental preparation symbolized by $|\psi\rangle$[35]. This evolution is not accessible to the senses, since its observation would consist in performing one or several measurements which would immediately stop it (Bitbol 1996)[36]. With an ontology of mere experimental contexts and measurement outcomes, there is no question of cat, as a "substantial and permanent object"[37], and there is consequently no question of its state or properties, hence "ridiculous feline situations" (in the words of Mackintosh 2019) are ruled out.

## 4.3 Consequence for the Einstein-Podolsky-Rosen (EPR) paradox

Let us now consider the paradox raised by Einstein, Podolsky and Rosen (1935), or EPR paradox. It relies upon the idea that quantum physics deals with physical systems (atoms, electrons, photons, …)[38]. The authors then deduce that quantum physics is unable to describe completely those systems (by attributing them, for instance, both a precise position and a precise momentum)[39], or that instantaneous signal transmission should occur between them, which would violate the theory of relativity (Bitbol 1996). With an ontology that has no room for any "'system' endowed with 'properties'… as in the good old days of classical physics" (in the words of Bitbol 2007), the EPR paradox obviously vanishes, similarly to Schrödinger's cat. With an ontology of mere experimental contexts and measurement outcomes, the only type of instantaneous event that occurs, besides a measurement result, is the change of experimental context, occurring precisely at the time when that result is obtained. The state vector $|\psi\rangle$ that characterizes the experimental preparation preceding the recording of value $a_i$ for the measured variable $A$, is instantaneously replaced by a new state vector $|\psi'\rangle$, at the very moment when this value $a_i$ is obtained[40], which value contributes to defining the new state vector $|\psi'\rangle$[41]. That is to say, this measurement leads immediately to a new experimental context, which includes the experimental preparation characterized by the new state vector $|\psi'\rangle$[42], but absolutely no signal is transmitted instantly (with the ontology derived above, by the way, one wonders between which entities such a signal would be transmitted). It is precisely that immediate change of experimental context that is interpreted as an instantaneous superluminal signal transmission, when quantum theory is referred to physical systems—however remote from each other they may be. This change, which then seems to result from the measurement of a so-called first system, modifies instantly the possible outcomes of the measurement of a so-called second system, said correlated to the first one[43].

---

[35] Nothing occurs "except the protean potentiality to occur". See section 4.1 above.

[36] As noted by Heisenberg, "while the theory determines what can be observed, … a theory also determines what cannot be observed". See (Holton 2000).

[37] In the words of Piaget (1954), according to whom this "… object concept, far from being innate or given ready-made in experience, is constructed little by little", and that, in six stages during the first two years after birth.

## 4.4 Consequence for the quantum measurement problem

In the following, I will consider the quantum measurement problem, of which the Schrödinger's cat thought—or Gedanken—experiment is in fact a mere illustration. More specifically, I will consider the so-called predictive version of the measurement problem (Bitbol 1996).

Let us assume that a measurement is performed using an apparatus denoted by $M_A$, following an experimental preparation said of the *first order*. The probability that a given measurement result is obtained at a given point in time is calculated (via Born's rule) using the state vector $|\psi\rangle$ (where the Greek letter ψ is a lowercase psi), whose evolution is governed by Schrödinger equation. Let us now assume that a measurement is performed following an extended experimental preparation, which includes the first-order apparatus $M_A$. The predictive content of this *second-order* preparation is symbolized by the overall state vector $|\Psi\rangle$ (where Ψ is this time an uppercase psi). According to the universality of quantum laws, a measurement result can be considered as being obtained in the first-order experimental context only at the moment when a measurement result is obtained in the second-order context. In other words, the evolution of the state vector $|\psi\rangle$ can be considered as being suspended only at the moment when the evolution of the overall state vector $|\Psi\rangle$ is itself suspended, "... as if the occurrence of the first-order experimental events were conditional on the completion of a second-order experimental procedure; and as if, consequently, the first-order equipment did not possess determinations of its own, but only determinations depending on those of the second-order equipment."[44] That is obviously incompatible with the "pre-understood certainties upon which the experimenters rely in their works and in their speeches" (Bitbol 1996), certainties according to which the concept of measurement precisely relies upon the idea that an equipment possesses its own determinations, a measurement outcome corresponding to one of those determinations. In sum, there is a contradiction between those certainties and the conclusions that flow from quantum theory, hence the measurement problem in its predictive version. With a mere ontology of experimental contexts and measurement outcomes the contradiction disappears, as well as the measurement problem, since this ontology prevents any reference to physical systems, including of course those of the category of material bodies, such as measuring apparatus (even though they were cats!). Indeed, there is then obviously no question of their determinations.

---

[38] That is, the Aristotelian idea that physics is first and foremost mechanics. See section 4.1 above and (Jaeger 2017).

[39] It is "… impossible… to get the full ID card of the system" (Auffèves and Grangier 2016).

[40] Jaeger (2017) speaks of "the exceptional sort of quantum state change taking place during measurement".

[41] The value of the measured variable is thus "a piece of information that helps making other predictions" (Bitbol 1996). See also (Wigner 1963 and 1967). As Wigner (1963) puts it, although this may face us with "a strange dualism", "… state vectors… change in two ways. As a result of the passage of time, continuously, according to Schrödinger's time-dependent equation… The state vector changes, however, also discontinuously, according to the probability laws, if a measurement is carried out…"

[42] Within the ontological framework described by Auffèves and Grangier, which is somewhat close to the one of this paper (see section 5 below), "… a measurement amounts to define… a new context…" (Auffèves and Grangier 2016).

[43] Incidentally, we have at hand a very simple explanation of time irreversibility. The succession of different experimental contexts depends on events—namely measurement outcomes—whose probability of occurrence is not—at least not always—equal to 1, which would be required for time reversibility.

## 5 Philosophical implications of the derived quantum ontology

From the point of view defended here, quantum physics is a theory that simply predicts measurement outcomes depending on experimental contexts, the latter themselves depending on previous measurement outcomes. It might be alleged that this is a just an instrumentalist (or pragmatist) account of quantum physics, according to which this "... science is reduced to a collection of recipes that work..." (Thom 2016)—that is, to "a set of techniques for making predictions" (Maudlin 2019)—and remains silent about what this theory refers to, and even more so about the world[45]. Driven by the antirealist attitude, an instrumentalist will most likely appreciate to refer to experimental contexts and measurement outcomes only, considering quantum physics as a mere predictive theory with no descriptive value. Yet, it is considered here that (a) the evolving experimental contexts understood as constrained possibilities of future measurement outcomes and (b) the instantaneous measurement outcomes themselves *are* the objects of quantum theory and are *the only ones,* which objects are *described* by certain elements of quantum formalism ($|\psi\rangle$, $a_i$, ...). This paper indeed aims at answering "the question of the object of quantum mechanics", while the instrumentalist tends to keep quiet about it[46]. It aims at offering "images" to the quantum physicist (see Bitbol 1996), and even images (those of experimental contexts and measurement outcomes) with a true physical meaning, insofar as they can be related to familiar representations (see Fig. 2). In sum, this paper is clearly about the ontology of quantum physics—even though the term "ontology" is understood in a weak, non-metaphysical sense (Bitbol 1996)—and does not simply settle for its "operationality" (a word used by d'Espagnat 1997).

It remains true that the ontology presented here seems to accord very naturally with instrumentalism provided that the instrumentalist, while maintaining an antirealist attitude, agrees to refer to the elements of this ontology *as if* they actually composed the sensible world, but without adhering to this belief (Bitbol 1994). She/he then just needs to substitute experimental contexts (symbolized by $|\psi\rangle$, $\langle a_i|\psi\rangle$, and $\hat{H}$) and measurement outcomes (symbolized by $a_i$) for the first and only "objectification level" that measuring apparatus and other laboratory instruments represent for her/him (Bitbol 1996). The substitution is made easier by the fact that experimental contexts and measurement outcomes are related to the same immediate sense experiences as those instruments (Fig. 2). Such contexts and outcomes can then be regarded as mere *knowledge* instruments—i.e. mere "abstractions that help us to organize our thinking" (in the words of Mermin 2014)—to which the instrumentalist can refer without scruples, insofar as they prove fruitful in avoiding paradoxes and other interpretation issues, as discussed in sections 4.2 to 4.4 above. One may even wonder whether the fruitfulness of such an ontology displays any limitations. Indeed, the ability of the "structural core" of quantum theory to provide (probabilistic) predictions of *measurement outcomes* on the basis of given *experimental contexts* has no "known limitations to date" (Bitbol 1996). If the reference to experimental contexts and measurement outcomes should confront us with limitations, the latter should then logically coincide with the limitations of quantum physics in its ability to provide such predictions. One may also wonder whether, in the end, an instrumentalist can really do without ontology[47]. Even if the quantum physicist regards her/his theory as a first and foremost predictive one, she/he still refers to entities, namely (a) "what the predictions are about"[48] and (b) the situations on the basis of which those predictions are made, those two kinds of entities being equivalent to (a) measurement outcomes and (b) experimental contexts, respectively[49]. In other words, even when quantum physics seems to be reduced to its "theoretical skeleton", it is still somehow covered with an "interpretative flesh" (in the words of Bitbol 1996), thin though it may be. This "interpretative flesh"—namely the ontology defended in this paper—is particularly adapted to the interpretation of quantum physics, since it is reduced to the strict minimum and thereby follows as closely as possible the forms of its "theoretical skeleton". If I may borrow the words of Rovelli (1996), "my effort here is not to modify quantum mechanics to make it consistent with my view of the world, but to modify my view of the world to make it consistent with quantum mechanics."

[44] "The same reasoning could also be applied to the second-order procedure, which would refer to a third-order procedure, etc." (Bitbol 1996), so that the second-order equipement would not seem to possess its own determinations either, etc.

[45] In line with one of Niels Bohr's attitudes towards quantum theory. See (Bitbol 1996).

[46] According to Bitbol (1996), instrumentalism is an "... antirealist attitude which tends to block the referential projection..."

Finally, the ontology presented here is perfectly compatible with a more committed philosophical attitude. There is indeed no impediment to elevating its elements to the rank of really existing entities. There is no impediment to considering that—besides time—measurement outcomes and experimental contexts are what exists[50] and *everything* that exists. In sum, there is nothing to prevent from seeing them as the ultimate reality, i.e. the fundamental elements of the sensible world[51]. Then, there is no longer question of a mere "predictive symbolism" (Bitbol 1996), since this symbolism is used to *describe* the sensible world[52] and the processes taking place in that world, namely the continuous evolution—according to the

---

[47] It is questionable whether she/he can completely "block the referential projection", whether "the suspension of the descriptive judgment" (Bitbol 1996) that tends to characterize her/him can be total.

[48] "This by which they are tested" (Bitbol 1996).

[49] With regards to measurement outcomes, even when "inserted in a predictive scheme", "… the statement specifying the value of the measured variable… consists of a *descriptive* position-taking…", even though this position-taking is "occasional" (Bitbol 1996, emphasis added).

[50] With regards to experimental contexts, it is worth noting that Heisenberg rehabilitates in quantum physics Aristotle's idea that "… the possibility or rather the 'tendency' towards a course of events possesses itself a kind of reality…" (Heisenberg 1959 and, for the translation from German, Jaeger 2017).

[51] To use the words of Rudolf Carnap (1956), besides the "high degree of efficiency" that the ontology presented here can provide to quantum physics, and regardless of this efficiency, one can very well believe in the real existence of the elements of this ontology, that is to say, one can very well "give an affirmative answer" to "the external question" of "the ontological reality" of those "kinds of entities".

[52] I have underlined above (see footnote 49) the descriptive nature of measurement outcomes.

Schrödinger equation—of the potentialities symbolized by $|\psi\rangle$, and the instantaneous change of experimental contexts each time a measurement result is obtained[53]. This offers scientists an alternative to the instrumentalist agnosticism towards the true nature of the world. According to this alternative, the world is obviously not a world of objects in the usual sense of the word "object", but a world of both constrained "mere potentialities" (the experimental contexts)[54] and, so to speak, mere actualities (the measurement outcomes). Basically, this alternative corresponds to a combination of a "realism of the potentiality" and a "realism of the actuality" (in the words of Bitbol 2019).

Of course, the "upward continuity issue" (Bitbol 1996) must be addressed: from those elements regarded as fundamental, how does the appearance of a world made of "things outside us"[55] such as material bodies, which things, according to "the natural attitude of the mind"[56], seem to face the observing subjects with whom we identify and to exist independently of them, how does this appearance arise (Fig. 3)[57]? However, this very sensitive issue will not be addressed in this paper, whose aim is primarily to present an ontology of quantum physics that is as much as possible "the exact replica of the theory's symbols" (in the words of Bitbol 1996). The fact is, when one refers to this ontology, one is lead to "… the conclusion that quantum mechanics only gives probability connections between subsequent observations…", a conclusion that perfectly accords with "the standard theory of measurement in quantum mechanics" (see Wigner 1963).

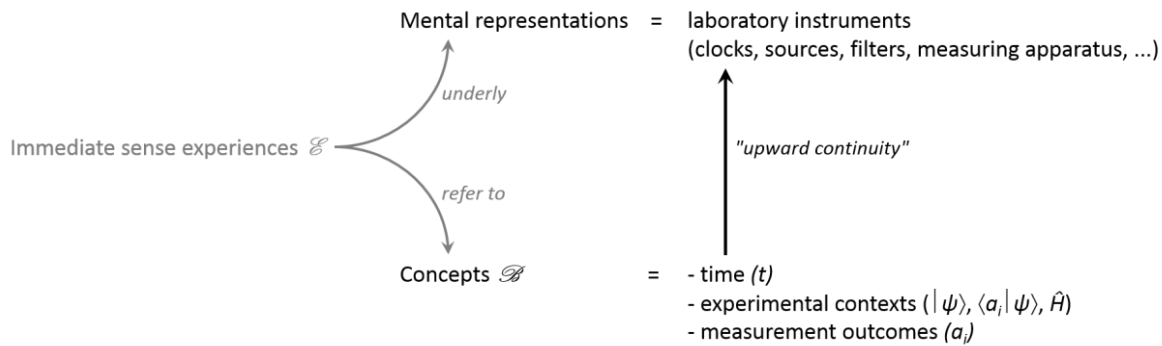

**Fig. 3** The ontology of evolving experimental contexts and instantaneous measurement outcomes, and the "upward continuity issue"

It should be noted out that compared to "the traditional ontology of material bodies" (Bitbol 1996), the difference is more radical with the ontology presented here than with that defended by Alexia Auffèves and Philippe Grangier. Indeed, besides experimental contexts, the latter ontology *still* involves physical systems, according to a conception close to the one adopted by Niels Bohr in response to the EPR paradox. According to Auffèves and Grangier

---

[53] "… In quantum theory, all physical processes are continuous and deterministic, except for observations, which are proclaimed to be instantaneous and probabilistic" (Proietti et al. 2019). See also (Everett 1957).

[54] See section 3 above, especially footnote 23.

[55] *Dinge außer uns* (Kant 1956).

[56] *Natürliche Geisteshaltung* (Husserl 1986).

[57] To use the words of Bitbol (1996, emphasis his), "… the proponents of an ontological revision have to show the *possibility* of an *upward* continuity from the object of quantum mechanics as they conceive it to the structure of the everyday environment." Indeed, the possibility for the revised ontology "… to be related to the categories of everyday language and action is… a crucial test for its universal applicability, and perhaps even for its viability."

(2016), the physical properties, renamed for the occasion "modalities", and corresponding to "measurement results", "… should not any more be considered as properties of the system itself, but jointly attributed to the system, and to the context in which it is embedded…"As for Bohr (1935, emphasis his), he writes that "… *the very conditions which define the possible types of predictions regarding the future behavior of the system*… constitute an inherent element of the description of any phenomenon to which the term 'physical reality' can be properly attached…"

Most likely because it is particularly parsimonious, the ontology defended in this paper appears as a kind of middle way par excellence. Apart from its compatibility with instrumentalism as well as with a more committed philosophical posture[58], this ontology is also compatible with both (a) dualism, opposing observing subjects to objects of perception conceived as outside those subjects and independent of them[59], and (b) monism—at least "neutral monism"—which overcomes this opposition, and whose relevance is highlighted by some authors (e.g. Bitbol 2019)[60]. The ontology defended by Auffèves and Grangier is incompatible with monism, as it aims at saving the dualist perspective of "physical realism", defined as follows: "… the goal of physics is to study *entities of the natural world, existing independently from any particular observer's perception,* and obeying universal and intelligible rules" (Auffèves and Grangier 2016, emphasis added).

## 6 Quantum physics as a theory of general relativity of experimental context

Besides its above-mentioned benefits, the quantum ontology derived in this paper seems to have great merit. By limiting the objects of this science to instantaneous measurement outcomes and—possibly—evolving experimental contexts, it indeed sheds light on what quantum physics may ultimately be, namely a theory of general relativity of experimental context. I shall explain this by focusing on the universality of quantum formalism.

According to this universality, the validity of quantum formalism shows "indifference to specific situations" (in the words of Bitbol 1996). In other words, equations (1) and (2) remain valid whatever the preparation of a physical measurement, that is, whatever the experimental context. In particular, equations (1) and (2) are valid for any so-called first-order *experimental context*[61]. For this reason, quantum physics can be identified with a theory of *special* relativity of experimental context. Equations (1) and (2) are also valid for any experimental context of higher order[62]. That is why quantum physics can be identified with a theory of *general* relativity of experimental context, according to which *all* experimental contexts are equivalent in regard to quantum laws.

Quantum physics can well be qualified as a relativity theory as understood by Einstein, namely a theory according to which physical laws "hold good" whatever the situation: in the case of the theory developed by the German-born physicist, which is basically a theory of the relativity of motion, the laws considered are valid "for all frames of references" (Einstein 1905a), that is, in all *spatio-temporal* situations; in the case of quantum physics, the laws considered are valid in all *experimental* situations[63]. Further, quantum theory can well be qualified as a theory of general relativity, insofar as the validity of the quantum laws is not restricted to first-order experimental contexts, but is *generalized* to contexts including themselves first-order contexts and so forth, i.e. to any *metacontext,* whatever its degree of complexity. A metacontext is then to a first-order context what, in Einstein's theory of relativity, acceleration is to rectilinear uniform motion.

---

[58] Apart from its ability, on the one hand, to be perfectly consistent with the idea that "the minimal purpose of a physical theory consists in providing predictions about measurement outcomes, on the basis of a given preparation" (Bitbol 1996) and, on the other hand, to serve the more ambitious objective of telling us something about the sensible world.

[59] As Auffèves and Grangier (2016), one will then consider experimental contexts as outside—preexisting—observing subjects, and one will do so for measurement outcomes.

[60] Experimental contexts and measurement outcomes will then be considered as that from which emerges the "dualist fiction" (Bitbol 2019), in which case the "upward continuity issue" might translate into the issue of the link between different theories of knowledge rather than between different ontologies (Batard, in preparation). Of note, "Russell repeatedly stated that neutral monism definitively solves the mind-body problem" (Woodfield 1999).

[61] See section 4.4 above about the notion of first-order context.

[62] See again section 4.4 above about the notions of second- and higher-order contexts.

To support the idea that quantum physics is a theory of general relativity of experimental context, let us come back to the quantum measurement problem as presented in section 4.4. One can very well imagine that, at a given time, the evolution of the state vector contributing to the characterization of a given context (let us say, a first-order context) is suspended *from the point of view of that context,* but at the same time it is still not suspended *from the point of view of a higher-order context* (say, a second-order context). In other words, one can very well imagine that a measurement result is obtained in a given context, while at the same time no measurement result is obtained in that context *considered from the point of view of a higher-order context*—in the same way that, according to Einstein's relativity theory, "… two events which, viewed from a system of co-ordinates, are simultaneous, can no longer be looked upon as simultaneous events when envisaged from a system which is in motion relatively to that system" (Einstein 1905a). What would be true—or, to be more precise, what would be *actual*— in a given experimental context would not be so in another, as well as, according to the relativity theory, what is true for an observer in a moving wagon (for instance the simultaneity of perception of two light signals coming from both ends of the wagon) is not true for someone on the platform. What is imagined here corresponds in fact to nothing else but the so-called "Wigner's friend paradox".

In the Gedankenexperiment devised by Nobel laureate Eugene Wigner, a friend of his makes an observation in a given experimental context, and Wigner himself inquires about the whole situation. The latter is described by an overall state vector that is still continuously evolving after the friend has made his observation, and keeps evolving so until the friend tells Wigner what precisely he has observed. Only then—that is, possibly long after the friend's observation—does a change in the overall state vector occur. Thus, from Wigner's point of view, no observation was made until his friend tells him *the outcome* thereof, for instance "seeing a flash" or "not seeing a flash". "However," Wigner comments, "if after having completed the whole experiment I ask my friend, 'What did you feel about the flash before I asked you?' he will answer, 'I told you already, I did [did not] see a flash,' as the case may be. In other words, the question whether he did or did not see the flash was already decided in his mind, before I asked him" (Wigner 1967). Consequently, for a while—however long that may be—there is a "contradiction" [*sic*] between what is a fact from the "friend's" point of view— say, he sees or has seen a flash—and what is a fact from "Wigner's" one—a fact "… from which he concludes that his friend cannot have recorded a definite outcome", so that all this time the two "… observers irreconcilably disagree about what happened in an experiment" (Proietti et al. 2019)[64]. Hence the thought experiment is called a paradox. To solve the paradox, Wigner argues that "observers with a consciousness" violate quantum laws, contrary to an "inanimate measuring device"[65]. According to the Hungarian-born physicist the situation of such a device—e.g. "an atom"—is *properly* described from "Wigner's" point of view, that is to say, there is no such thing as an observation made by the device as long as "Wigner" himself, "as ultimate observer", has not observed the whole situation—which is obviously not the case "if the atom is replaced by a conscious being" (Wigner 1967). Nevertheless, this claim does not seem to solve the paradox. Indeed, recent experimental results obtained from "an extended Wigner's friend scenario", where the conscious "friend" and "Wigner" himself are substituted by such inanimate devices, lead to the same conclusions as Wigner's original thought experiment. According to those results, the facts recorded by a device are in conflict with the ones recorded by an outside observer, hence the authors' conclusion that "facts of the world" are not "observer-independent", even if the observer is inanimate (Proietti et al. 2019)[66].

[63] Situations—or contexts—which, let us recall, include the variable to be measured. See section 3 above.

[64] Surprisingly enough, "the friend can even tell Wigner that [he] recorded a definite outcome *(without revealing the result),* yet Wigner and his friend's respective descriptions remain unchanged" (Proietti et al. 2019, emphasis added).

[65] "… The consciousness modifies the usual laws of physics", he writes (Wigner 1967).

If quantum physics is considered a theory of general relativity of experimental context, meaning that the occurrence—or actuality—of a measurement result depends on the experimental context, not only is there no paradox at all, but a conflict between experimental facts, such as between "Wigner's" facts and his "friend's" facts, is fully expected—no matter whether the so-called "Wigner's friend" is an inanimate device or a conscious being—since "Wigner's" experimental context is different from his "friend's", hence the difference between their recorded facts. Such a conflict is obviously incompatible with the idea that quantum physics deals with physical systems endowed with intrinsic properties, insofar as such properties (e.g. a given position or a given momentum) are assumed to be observer-independent, and therefore independent of the context of their observations. Conversely, an ontology with no physical systems, as the one derived in this paper, accords perfectly with Wigner's friend scenarios in particular, and in general with the idea that quantum physics is a theory of relativity of experimental context. Of note, this idea that "quantum theory is to be regarded as a relativity theory" is also formulated by Martin Davis (1977), who suggests that quantum theory involves "the principle of relativity of quantum measurements", according to which *"all measurements are relative to the frame of observation"* (emphasis his). And as Michel Bitbol (2018 and 2019) points out, since quantum physics is a theory "… in which the determinations one wishes to anticipate are related to the context of their occurrence…", it "… could rightfully be called a 'general theory of correlational prediction'…"

To be sure, one can accord with Wigner's friend situations while still referring to physical systems. For that, one just needs to forsake the notion of systems' *intrinsic* properties, as proposed by Carlo Rovelli in his relational interpretation of quantum physics, or RQM. As Rovelli puts it, "the unease with the Lorentz transformations derived from a conceptual scheme in which an *incorrect notion*—absolute simultaneity—was assumed, yielding all sorts of paradoxical situations. Once this notion was removed, the physical interpretation of the Lorentz transformations stood clear…" Drawing a parallel with that circumstance, he considers "… the hypothesis that all 'paradoxical' situations associated with quantum mechanics—such as the famous and unfortunate half-dead Schrödinger cat…—may derive from some analogous *incorrect notion* that we use in thinking about quantum mechanics." He then suggests that such an incorrect notion is that of "absolute, or observer-independent, state of a system" (Rovelli 1996, emphasis his). In other words, "the quantum state of a system is always a state of that system *with respect to a certain other system*"—the latter being "considered as an observer"[67]—so that there is no "absolute properties that the system has at a certain time", i.e. no intrinsic system properties (Rovelli 1996, emphasis added)[68]. Yet, one may wonder why still referring to physical systems if they no longer carry observable intrinsic properties[69]. What is the ultimate rationale of such a reference, apart from a last—and desparate?—attempt to somehow retain the everyday life concept of material bodies—e.g. through the "cherished ontology of elementary particles" (in the words of Quine 1990)—whereas, clearly, "... the traditional ontology of material bodies endowed with properties appears to be at odds with quantum mechanics as well as relativistic mechanics" (Bitbol 1996)? Feynman (2006, emphasis added) points out that "the theory of quantum electrodynamics describes Nature as absurd *from the point of view of common sense.*" But is it not precisely because common sense requires that quantum theory refers to something—namely physical systems—that reminds us of materials bodies in some way? Is it not because, so to speak, the common notion of material bodies pollutes the mind of the scientist, the concept of physical systems being a vestigial remnant of such contamination? If one is to do away with the notion of intrinsic properties, why not be ontologically even more parsimonious[70]? Certainly, Rovelli (1996) takes "to some extreme consequences" the lesson of Heisenberg according to which "... we should stop thinking of the 'state of the system'...", meaning: the *intrinsic* or *absolute* state of the system. By presenting a quantum ontology devoid of any physical system, I somehow take Heisenberg's lesson to even more extreme consequences. While it is meaningless to refer to the intrinsic state of a system in RQM, the quantum ontology presented here renders meaningless the reference to the very notion of system. Interestingly enough, Rovelli (1996, emphasis added) suggests that the "... incorrect notion that generates the unease with quantum mechanics is the notion of 'observer-independent state' of a system, *or 'observer-independent values of physical quantities'.*" Let us consider the words in italics. After all, nothing imposes that physical quantities are to be referred to physical systems. The notion of "observer-independent state of a system" and that of "observer-independent values of physical quantities" are therefore not equivalent, contrary to the claim of Rovelli (1996). One can then envisage that when it comes to interpreting quantum theory, the incorrect notion is merely that of system itself, and that quantum physics is *not* "a theory about the physical description of physical systems"—contrary to the idea of the Italian-born physicist (see again Rovelli 1996)—that is to say, it is not mechanics (see section 4.1 above).

Quantum Bayesianism, or QBism, is an alternate view of quantum physics that accords with Wigner's friend situations. Among other claims, a QBist states the following (Fuchs et al. 2014): (a) "... quantum mechanics is a tool anyone can use to evaluate, on the basis of one's past experience, one's probabilistic expectations for one's subsequent experience", hence "the personal character of probability" in question[71]; (b) a quantum measurement outcome is such an experience, which is also personal—i.e. eminently private—and therefore observer-dependent. "The disagreement between Wigner's account and his friend's is paradoxical", Fuchs et al. (2014) explain, "only if you take a measurement outcome to be an objective feature of the world, rather than the contents of an agent's experience. The paradox vanishes with the characterized with respect to other physical systems", since it turns out that "... the traditional concept of properties which belong to the system is incompatible with quantum mechanics."

---

[69] As just seen, Rovelli's reference to systems is unambiguous and even necessary, since in RQM the notions of state and property are "other system"-dependent. "I assume that the world can be decomposed (possibly in a large number of ways) into a collection of systems...", Rovelli (1996) further writes.

[70] "The concept that quantum mechanics forces us to give up", Rovelli (1996) states, "is the concept of a description of a system independent of the observer providing such a description; that is, the concept of absolute state of a system." One might be tempted to ask him "Why not the concept of system itself?" He goes on to write, "The structure of the classical scientific description of the world in terms of *systems* that are in certain *states* is perhaps incorrect, and inappropriate to describe the world beyond the $\hbar \to 0$ limit" (emphasis his). One might then want to ask him "Why abandon the concept of state and keep that of system?"

[71] "A QBist takes quantum mechanics to be a personal mode of thought—a very powerful tool that any agent can use to organize her own experience", Fuchs et al. (2014) further state.

recognition that a measurement outcome is personal to the experiencing agent. There is an outcome in the friend's experience; there is none yet in Wigner's. Of course their accounts differ", and more generally "… reality differs from one agent to another", but the accounts they give are equally correct.

The QBist idea that measurement outcomes are singular—nonuniversal—entities is fully compatible with the ontology derived in the present paper. In fact, the latter even implies such an idea. When quantum theory is interpreted on the ground of that ontology, measurement outcomes are always entirely dependent on experimental contexts[72], which are themselves singular—nonuniversal—entities, as seen above with Wigner's friend[73]. In return, however, QBism does not accord with the ontology I defend here, since QBists unambiguously keep referring to physical systems, "whether they be atoms, enormous molecules, macroscopic crystals, …" (Fuchs et al. 2014). That being said, let us consider as QBists seem to do that the only things one can ultimately access are one's "own *or* individual *or* private *or* personal experiences"[74], i.e. one's "*internal* personal experiences"[75]. One may then wonder what motivates QBists to relate those internal experiences to so-called "*external* (physical) systems" such as "particles" (Fuchs et al. 2014, emphasis added)[76], apart from—once again—somehow safeguarding the everyday notion of "things outside us" such as material bodies. Would they claim that physical systems are mere mental constructs grounded in intimate experiences[77]? One could answer that such mental constructs have little to do with quantum physics.

As Fuchs et al. (2014) themselves rightly say, "given the unprecedented murkiness that has enveloped quantum foundations for the past nine decades it is not surprising that an unorthodox approach is needed to dispel the fog." It is however questionable whether QBism, by somehow retaining the notion of physical system, is unorthodox enough to do so[78]. Bitbol (2019) points out that during a radical scientific revolution, as the quantum revolution obviously is, "… it is the objects of research that can change completely, and not just be better characterized. In particular, one must recognize the fragility, contingency, and constituted character of the most traditional objects of physics, those that are so deeply engrained in its tradition that renouncing them seems to amount to renouncing physics itself: the things-substances, bodies and subsisting material points." In view of the physical ontology derived in this paper and its ability to solve—or rather dissolve—well-known paradoxes and interpretation problems, as well as to reveal the possible ultimate meaning of quantum theory, the quantum physicist is invited to renounce the general notion of physical system, radical though this renunciation may be.

---

[72] "Contextual phenomena are *contingent* in respect to classical laws but *necessary* in regard to quantum laws" (Pringe 2007, emphasis his).

[73] This is why, by the way, that ontology forbids any "description of the universe *'in toto'*", as also does RQM (Rovelli 1996). In this regard, Kant interestingly points out that "the universe as a whole is not an object of possible experience" (Winterbourne 2012; see Kant 1956).

[74] "What is real for an agent rests entirely on what that agent experiences…", Fuchs et al. (2014) write.

[75] "A measurement outcome… is created for the agent who takes the measurement action only when it *enters* the experience of that agent" (Fuchs et al. 2014, emphasis added).

[76] "In QBism… an agent applies quantum mechanics to *everything* outside her internal personal experience" (Fuchs et al. 2014, emphasis theirs).

[77] "… Any user's own experience constitutes all of the raw material out of which she constructs her world", Fuchs et al. (2014) state.

[78] As a matter of fact, "… QBism is not all that radical", Fuchs et al. (2014) admit.

# 7 Conclusion

What is quantum physics about? What is the meaning of this theory? In this paper, an ontology of quantum physics—i.e. what it may ultimately be *about*—has been derived from two comments by Albert Einstein related to his general approach to physical theories. Entities of this ontology are strictly limited to evolving experimental contexts and instantaneous measurement outcomes. Those two kinds of entities are to be understood, respectively, as constrained "pure potentialities"—namely, mere *potential* measurement outcomes—and, so to speak, "pure actualities"—namely, mere *actual* measurement outcomes—, the potentialities changing each time an actuality—i.e. a measurement event—occurs. Material bodies in particular and physical systems in general (e.g. particles such as atoms, electrons, photons) are clearly excluded from the ontology in question[79]. Hence the latter can be referred to as an "anti-system" ontology, and quantum physics should no longer be identified with mechanics.

"As in the good old days of classical physics", most—if not all—physicists use to talk about isolating a system so as to measure its properties (e.g. Auffèves and Grangier 2016), whether they be intrinsic properties or observer-dependent ones. Within the thought framework derived in this paper, this so-called isolation merely amounts to contributing to the definition of a given experimental context, i.e. to the definition of potential measurement results, waiting for such a result to occur.

Besides its ability to clear up well-known conundrums of quantum physics, that "anti-system" ontology sheds light on what quantum theory may fundamentally be—i.e. its ultimate *meaning*—namely a theory of general relativity of experimental context. Indeed, quantum laws are laws of probabilistic predictions of measurement outcomes[80] that are valid in any experimental context, that is, regardless of the corresponding size (whether one is dealing with a mere darkroom or a particle accelerator) and complexity (be it a first-order context or a metacontext). In view of that, and as illustrated by Wigner's friend paradox, quantum physics implies an alternative conception of objectivity in the sciences of nature. This conception corresponds to the generalization of the principle of relativity, whereby the laws of nature—whatsoever—hold good whatever the situation. According to this alternate conception, objectivity is not related to the observation of the same *things* (e.g. physical systems, such as material bodies) or even *facts* (e.g. measurement outcomes) by independent observers, but to the recorded obedience to the same *laws,* on the basis of observations obtained in different experimental contexts, however much those observations may be in conflict with one another. In sum, there is no longer question of universal things or even universal facts, but only of universal laws of the sensible world.

"This world, the same for all…" says the Presocratic Greek philosopher Heraclitus of Ephesus[81]. It is the same for all *merely in regard to its laws,* in keeping with the meaning of the Greek word "kosmos" (κόσμος), implying the notion of order.

**Acknowledgements**     I am very grateful to Michel Bitbol (CNRS, Archives Husserl, École Normale Supérieure, Paris) and Vincent Vennin (CNRS, Laboratoire AstroParticules et Cosmologie, Paris) for their valuable bibliographical suggestions and their fruitful and encouraging feedbacks on this paper.

---

[79] This obviously contrasts with the "desiderata for an interpretation of quantum mechanics" of Mermin (1998) and most probably many physicists.

[80] Quantum "… laws of nature do not determine the occurrence of an event, but the probability of this occurrence" (Heisenberg 1959). In other words, those "… laws govern the possible and not the actual" (Heisenberg 1955). See footnote 30 above.

[81] "Kosmon tonde, ton auton hapantôn…" (κόσμον τόνδε, τὸν αὐτὸν ἁπάντων…). See (Héraclite 1986) and (Brague 2003).

**Thierry Batard** received a BA in Philosophy from Reims University (France), a MA in Philosophy and History of Sciences from Paris 1 University (Panthéon-Sorbonne), and an engineering degree and a PhD from AgroParisTech.